\documentclass[aps,prc,preprint,superscriptaddress]{revtex4-1}
\usepackage[pdftex]{graphicx}
\usepackage{color}

\begin{document}


\title{New $\alpha$ ANC measurement of the 1/2$^+$ state in $^{17}$O at 6.356 MeV, that dominates 
the $^{13}$C($\alpha$,n)$^{16}$O reaction rate at temperatures relevant for the s-process}


\author{M.L. Avila}
\email[]{mavila@anl.gov}

\altaffiliation{Physics Division, Argonne National Laboratory, Argonne IL 60439, USA}
\affiliation{Department of Physics, Florida State University, Tallahassee, FL 32306, USA}

\author{G.V. Rogachev}
\email{rogachev@tamu.edu}
\affiliation{Department of Physics\&Astronomy and Cyclotron Institute, Texas A\&M University, College Station, 
TX 77843, USA}

\author{E. Koshchiy}
\affiliation{Department of Physics\&Astronomy and Cyclotron Institute, Texas A\&M University, College Station, 
TX 77843, USA}

\author{L.T. Baby}
\affiliation{Department of Physics, Florida State University, Tallahassee, FL 32306, USA}

\author{J. Belarge}
\affiliation{Department of Physics, Florida State University, Tallahassee, FL 32306, USA}

\author{K.W. Kemper}
\affiliation{Department of Physics, Florida State University, Tallahassee, FL 32306, USA}

\author{A.N. Kuchera}
\altaffiliation{National Superconducting Cyclotron Laboratory, Michigan State University, East Lansing, 
MI 48824, USA}
\affiliation{Department of Physics, Florida State University, Tallahassee, FL 32306, USA}

\author{D. Santiago-Gonzalez}
\altaffiliation{Department of Physics and Astronomy, Louisiana State University, Baton Rouge, LA 70803, USA}
\affiliation{Department of Physics, Florida State University, Tallahassee, FL 32306, USA}

\date{\today}

\begin{abstract}
\begin{description}
\item[Background]
Accurate knowledge of the $^{13}$C($\alpha$,$n$)$^{16}$O reaction cross section is important for the understanding of the
s-process in AGB stars, since it is considered to be the main source of neutrons. The sub-threshold 1/2$^+$ state at 
excitation energy of 6.356 MeV in $^{17}$O has a strong influence on the reaction cross section at energies relevant for astrophysics. 
Several experiments have been performed to determine the contribution of this state to the $^{13}$C($\alpha,n)^{16}$O reaction rate.
Nevertheless, significant discrepancies between different measurements remain.
\item[Purpose] The aim of this work is to investigate these discrepancies.
\item[Method] An 8 MeV $^{13}$C beam (below the Coulomb barrier) was used to study the $\alpha$-transfer reaction
$^6$Li($^{13}$C,$d$)$^{17}$O.
\item[Results] The squared Coulomb modified ANC of the 1/2$^+$ state in $^{17}$O measured in this work is
$(\tilde C^{^{17}\text{O}(1/2+)}_{\alpha-^{13}\text{C}})^2=3.6\pm0.7\hspace{0.2cm}\text{fm}^{-1}$.
\item[Conclusions] Discrepancy between the results of $\alpha$-transfer experiments have been resolved. However, 
some discrepancy with the most recent measurement using the Trojan Horse method remains.    
\end{description}
\end{abstract}


\maketitle

\section{Introduction}
The slow neutron capture process or s-process, occurs in a relatively low neutron
density environment in Asymptotic Giant Branch (AGB) stars. 
This process is essential for the nucleosynthesis of heavier
elements. It is believed that the s-process is responsible for
nearly half of the heavy elements observed in the Universe \cite{Came55}.
 The main characteristic of this process,  is that the neutron
capture is slower than the $\beta$ decay.
At low temperature ($<$ 10$^8$ K) for low mass stars the 
$^{13}$C($\alpha,n$)$^{16}$O reaction plays the major role and is considered to be
the main source of neutrons for the s-process in such stars \cite{Iben83}.
Thus, this reaction rate is a necessary ingredient for constraining the models 
of AGB stars. Direct measurements are only available for center of mass energies above
279 keV. Below this energy the cross section has to be extrapolated. 
Extrapolation to lower energies causes a large
uncertainly due to the presence of a sub-threshold 1/2$^+$ resonance in
$^{17}$O at excitation energy of 6.356 MeV (3 keV below the $^{13}$C+$\alpha$
threshold energy).
This sub-threshold resonance enhances the 
cross section at low energies making an important contribution to 
the astrophysical S-factor.

The 1/2$^+$ state at 6.356 MeV has been the subject of several investigations
\cite{Kubo03,Keel03,John06,Pell08,Heil08,Guo12,LaCo13}. \citet{Kubo03} measured the spectroscopic factor of the 1/2$^+$ 
state using the $^{13}$C($^6$Li,$d$)$^{17}$O reaction at 60 MeV of $^6$Li. A very small $\alpha$ spectroscopic factor
of $S_{\alpha}=0.011$ was found in this study. This value suggested a very
small contribution of this sub-threshold state to the cross section. 
However, using the same experimental data it was shown by \citet{Keel03} that the data
is consistent with $S_{\alpha}$ ranging form 0.15 to 0.41 depending on the DWBA 
parameters used. Later, \citet{John06} studied the $\alpha$-transfer reaction $^{13}$C($^6$Li,$d$)$^{17}$O at 
sub-Coulomb energies obtaining a squared Coulomb modified ANC value of 
$(\tilde C^{^{17}\text{O}(1/2+)}_{\alpha-^{13}\text{C}})^2=0.89\pm0.23\hspace{0.2cm}\text{fm}^{-1}$. 
This ANC corresponds to a contribution of this state to 
the astrophysical S-factor at zero energy of a factor of five larger than the one extracted in Ref. \cite{Kubo03}. 
However it produced a relatively small enhancement at the Gamow window energies for the 
s-process (around 180 keV). \citet{Pell08} used the
$^{13}$C($^7$Li,$t$)$^{17}$O transfer reaction above Coulomb barrier energies ($^7$Li beam energies of 28 and 34 MeV)
obtaining a squared Coulomb modified ANC of
$(\tilde C^{^{17}\text{O}(1/2+)}_{\alpha-^{13}\text{C}})^2=4.5\pm2.2\hspace{0.2cm}\text{fm}^{-1}$,
which is a factor of 5 larger than the one from \citet{John06}. \citet{Heil08} performed a comprehensive R-Matrix analyses
of all available data, including the direct measurements 
of the $^{13}$C($\alpha,n$)$^{16}$O reaction cross section down to a c.m. energy of 320 keV and extrapolated the cross section to 
the astrophysically relevant energies (below 200 keV). Their extrapolations agree with the S-factor curve found by 
\citet{Pell08}. \citet{Guo12} evaluated a squared Coulomb modified ANC of 
$(\tilde C^{^{17}\text{O}(1/2+)}_{\alpha-^{13}\text{C}})^2=4.0\pm1.1\hspace{0.2cm}\text{fm}^{-1}$
using the reaction $^{13}$C($^{11}$B,$^7$Li)$^{17}$O with $^{11}$B beam energy of 50 MeV. This result is in good agreement
with \citet{Pell08}. More recently, the ANC of the 1/2$^+$ state was determined by \citet{LaCo12}
using the Trojan Horse technique ($^6$Li beam energy of 7.82 MeV) extracting a squared Coulomb modified ANC value of 
$(\tilde C^{^{17}\text{O}(1/2+)}_{\alpha-^{13}\text{C}})^2=6.7\pm0.6\hspace{0.2cm}$fm$^{-1}$ that was later revised
to $(\tilde C^{^{17}\text{O}(1/2+)}_{\alpha-^{13}\text{C}})^2=7.7\pm0.3_{\text{stat}}$$_{-1.5\hspace{0.05cm}
\text{norm}}^{+1.6}\hspace{0.2cm}$fm$^{-1}$ in Ref. \cite{LaCo13}. This value
is higher than those reported in Refs. \cite{Pell08,Guo12}, however it is within the large uncertainties
of Ref. \cite{Pell08}.

The values of the ANC measured at above barrier energies in Refs. \cite{Pell08,Guo12,LaCo13} are at least 4 times
larger than the value measured at the sub-Coulomb energy in Ref. \cite{John06}. The main goal of this work is to remeasure 
the ANC using the sub-Coulomb $\alpha$-transfer reaction used by
\citet{John06} and resolve the discrepancies between the ANC measured at sub-Coulomb energies and 
the ANC measured at above Coulomb barrier energies. The application of the sub-Coulomb $\alpha$-transfer ANC technique was 
pioneered in Ref. \cite{Brun99},
where the $\alpha$-transfer reactions $^{12}$C($^6$Li,$d$)$^{16}$O and $^{12}$C($^7$Li,$t$)$^{16}$O were used
to study the sub-threshold 2$^+$ and 1$^-$ states in $^{16}$O at 6.92 and 7.12 MeV, respectively.
More recently, the validity of the sub-Coulomb $\alpha$-transfer approach was demonstrated in Ref. \cite{Avil14}
by measuring the ANC of the 1$^-$ state at 5.9 MeV in $^{20}$Ne and comparing it to the well known width of this state.
Also, the 0$^+$ and 3$^-$ cascade transitions for one of the most astrophysically important reactions,
$^{12}$C($\alpha$,$\gamma$)$^{16}$O, were constrained using this approach \cite{Avila15}.

\section{Experimental setup and analysis}
The experiment was carried out at the John D. Fox Superconducting Accelerator Laboratory, at Florida State 
University. A $^{13}$C beam at energy of 8 MeV was used.
Inverse kinematics (heavy beam and light target) was chosen to achieve the lowest energies in the center of mass
(c.m.) and still be able to detect the recoil deuterons. The $^{13}$C beam was delivered by an FN Tandem Van de 
Graaff accelerator using a SNICS-II cesium-sputter ion source. Several different $^6$Li targets of about 
35 $\mu \text{g}/\text{cm}^2$ thick were used. The $^6$Li targets were prepared under vacuum and transported
to the chamber in a vacuum container to prevent oxidation. For the identification of the reaction 
products two $\Delta E$-$E$ telescopes were mounted on remotely controlled rotating rings placed to the right
and left of the beam axis. Each of the $\Delta E$-$E$ telescopes were constructed with four pin diode 
2$\times$2 cm$^2$ silicon detectors and one position sensitive proportional counter wire, contained in 
a box filled with a P10 gas (10\% methane and 90\% Ar gas mixture). 
A Kapton foil of 7.5 $\mu$m thickness was used as the entrance window 
separating the P10 gas inside the detector from the chamber vacuum. This setup allows the measurement and identification of 
deuterons down to an energy of 1.8 MeV when a pressure of 150 Torr for P10 is used and also to observe the 
$^6$Li ions scattered at forward angles in the laboratory reference frame when the 
pressure in the proportional counters is reduced to 50 Torr. The intensity of the incoming 
beam was measured using a Faraday cup placed at the end of the chamber. 

The target thickness was measured using elastically scattered $^6$Li from the target (at forward angles in the
laboratory reference frame) by a $^{13}$C beam at 8 MeV and a $^{16}$O beam
at 10 MeV. Two different beams were used to evaluate the systematic uncertainty of the target 
thickness measurements which turned out to be 10\%. Elastic scattering of $^6$Li was also used to monitor target
integrity and effective thickness. The control measurements were performed every time a new target was used and 
after about three hours of use.  It was found that after 3-5 hours of target usage the energy of the elastically 
scattered $^6$Li reduces slightly. This was attributed to some material buildup on the surface of the target (probably 
carbon or oxygen). Since the
sub-Coulomb $\alpha$-transfer cross section is very sensitive to the energy of the beam, the targets were changed
every 3 to 5 hours and corresponding corrections were implemented. Details are given in Refs. \cite{Avila13,Avil14}. 
After the energy correction due to the energy loss in the buildup material in the target, an effective energy
of interaction of 7.72 MeV was calculated by taking into account the energy dependence of the cross section.
The effective energy of interaction is calculated by integrating the cross section over the full target thickness
and obtaining the value of the energy in the target at wich one-half of the yield is achieve, as is explained in Ref.
\cite{Rolfs}. Thus, the beam energy of 7.72 MeV is used for all the calculations presented in this work.

\begin{figure}
\includegraphics[scale=0.43]{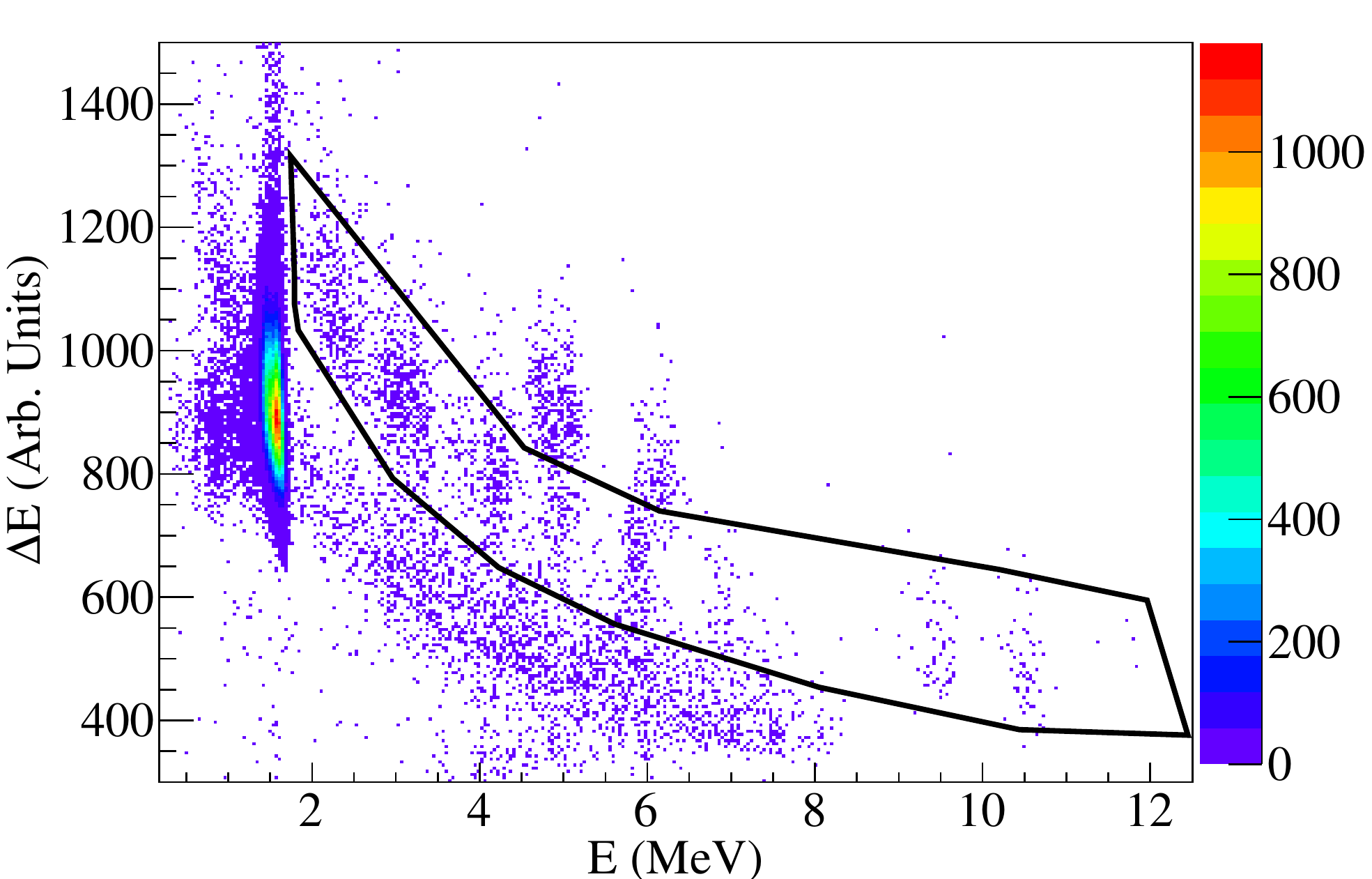}
\caption{\label{fig:EdE_17O_pin2} $\Delta E$ vs $E$ 2D scatter plot observed by a pin
detector at 20$^\circ$ in the laboratory reference frame and used for particle identification. 
The typical deuterons cut is shown by solid line. }
\end{figure}

\begin{figure}
\begin{center}\includegraphics[scale=0.45]{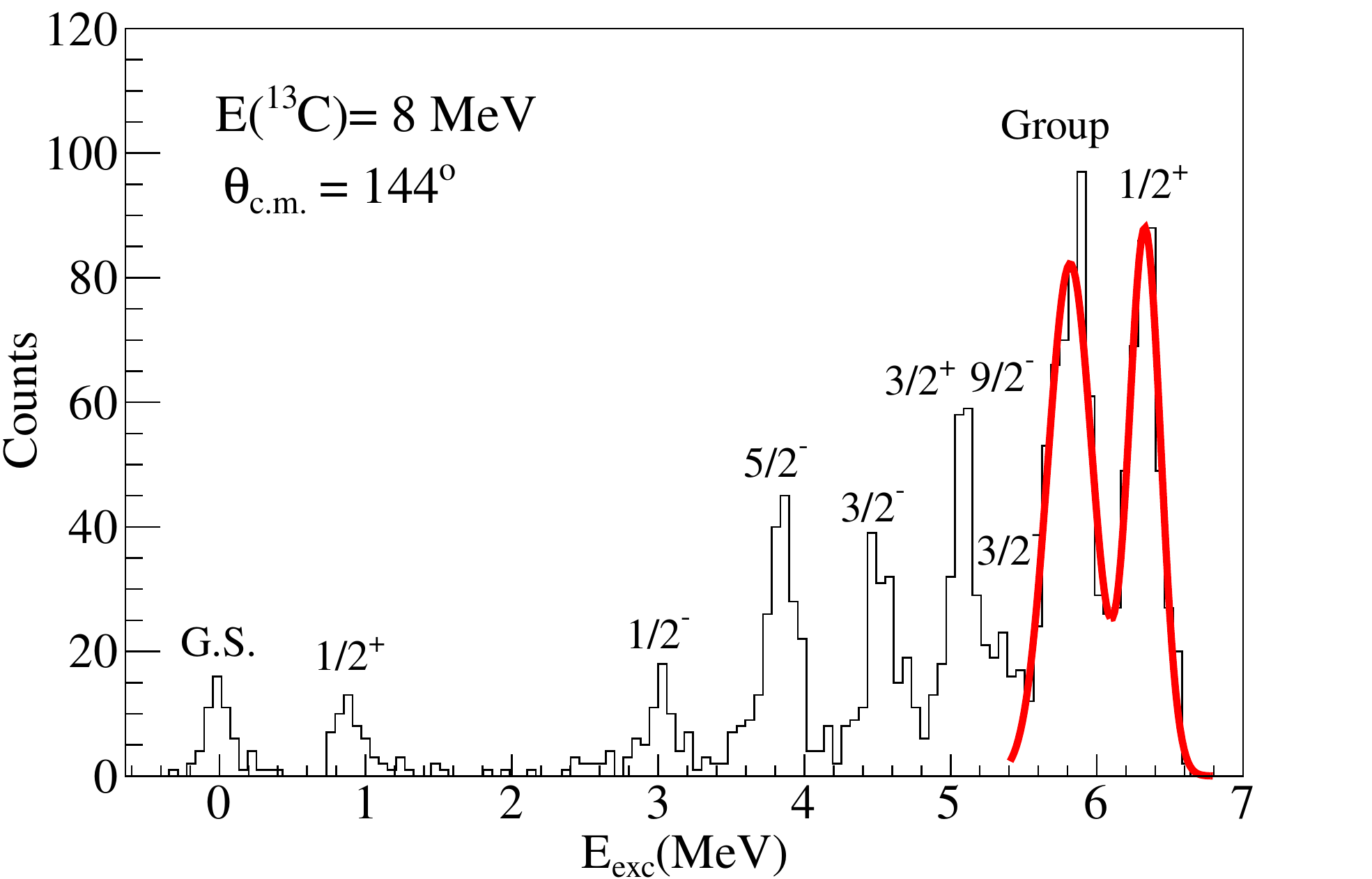}\end{center}
\caption{\label{fig:ExcE_17O_pin2} Spectrum of deuterons from the $^6$Li($^{13}$C$,d)^{17}$O reaction 
at 8 MeV (7.72 MeV effective energy after energy loss corrections)} of $^{13}$C beam at 144$^{\circ}$ in c.m. 
\end{figure}

A two-dimensional $E$ vs $\Delta E$ spectrum is shown in Fig. \ref{fig:EdE_17O_pin2} 
for a laboratory angle of 20$^\circ$. In this figure an intense proton region is observed at around 1.8 MeV. This proton
background corresponds to $^{13}$C+$p$ elastic scattering due to hydrogen contained in the target.
In Fig. \ref{fig:EdE_17O_pin2} the deuterons that correspond to the 1/2$^+$ state of interest are located
at the left corner of the deuteron contour. This state can be well separated from the intense proton background 
from the elastically scattered protons. The reconstructed excitation energy of $^{17}$O at the center of mass angle
of 144$^\circ$ is shown in Fig. \ref{fig:ExcE_17O_pin2}. In this figure all the labeled peaks that are produced 
correspond to well known states. The peak labeled as Group contains the four states 7/2$^+$(5.70 MeV), (5/2$^-$)(5.73 MeV),
3/2$^+$ (5.87 MeV) and 1/2$^-$ (5.94 MeV) that could not be resolved. It can be seen in this figure that the
1/2$^+$ state at 6.356 MeV is well separated from this neighboring group of states. 

\begin{figure}
\begin{center}\includegraphics[scale=0.44]{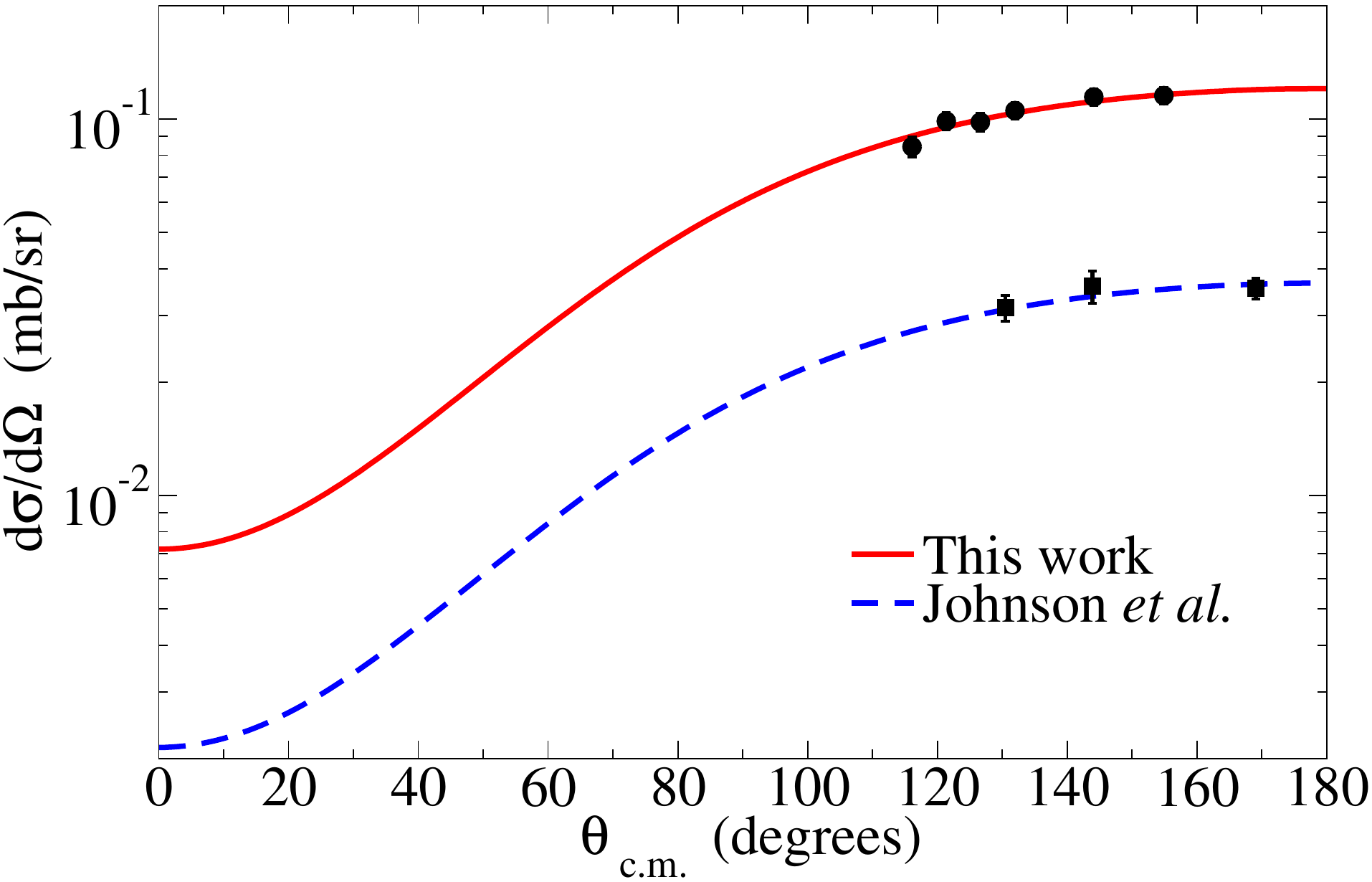}\end{center}
\caption{\label{fig:ang_dist_17O} Cross section and DWBA fit as a function
of center of mass angle of the 1/2$^+$ 
sub-threshold resonance state
of excitation energy 6.356 MeV in $^{17}$O measured in the present work (solid line) and in Ref. \cite{John06} (dashed line)
using a $^{13}$C beam energy of 7.72 MeV and 7.81 MeV respectively.}
\end{figure}

The angular distribution for population of the 1/2$^+$ state at 6.356 MeV and the Distorted Wave Born Approximation (DWBA)
fit are shown in Fig.
\ref{fig:ang_dist_17O}.  The DWBA calculations were done using the \textsc{fresco} 
code (version FRES 2.9) \cite{thom88} in the finite-range transfer approach with a full complex remnant
term. The $d$+$^{17}$O and $d$+$^{12}$C optical parameters that were used are the same 
parameters used to fit the $^{13}$C($^6$Li,$d$)$^{17}$O data in \cite{John06}.
The radius of $R_v=1.9$ fm and diffuseness of $a=0.65$ fm were used for the $\alpha$+$d$ form-factor potential. 
These parameters are the same as in \cite{Kubo72}. The potential depth was fitted to reproduce the 
binding energy of $^6$Li (1.474 MeV). We used three nodes (minimum+1) for the $\alpha$-cluster wave 
function of the 6.356 MeV 1/2$^+$ state in $^{17}$O (this excludes the node at the origin and infinity). 

The combined uncertainty of the squared ANC was determined from statistical uncertainty, uncertainties in the 
normalization procedure, and from uncertainties associated with the dependence of the results on the number of wave
functions nodes used and the optical potential parameters. If the potential parameters are kept within reasonable limits 
it was seen that the cross section varies by less than 15\%. In fact even if the nuclear part of the potentials are 
removed the results only
vary by about 30\%. The total uncertainty calculated for this ANC is 19\% and it is dominated by the optical potential parameters.

\section{Results}
\begin{figure}
\begin{center}\includegraphics[scale=0.43]{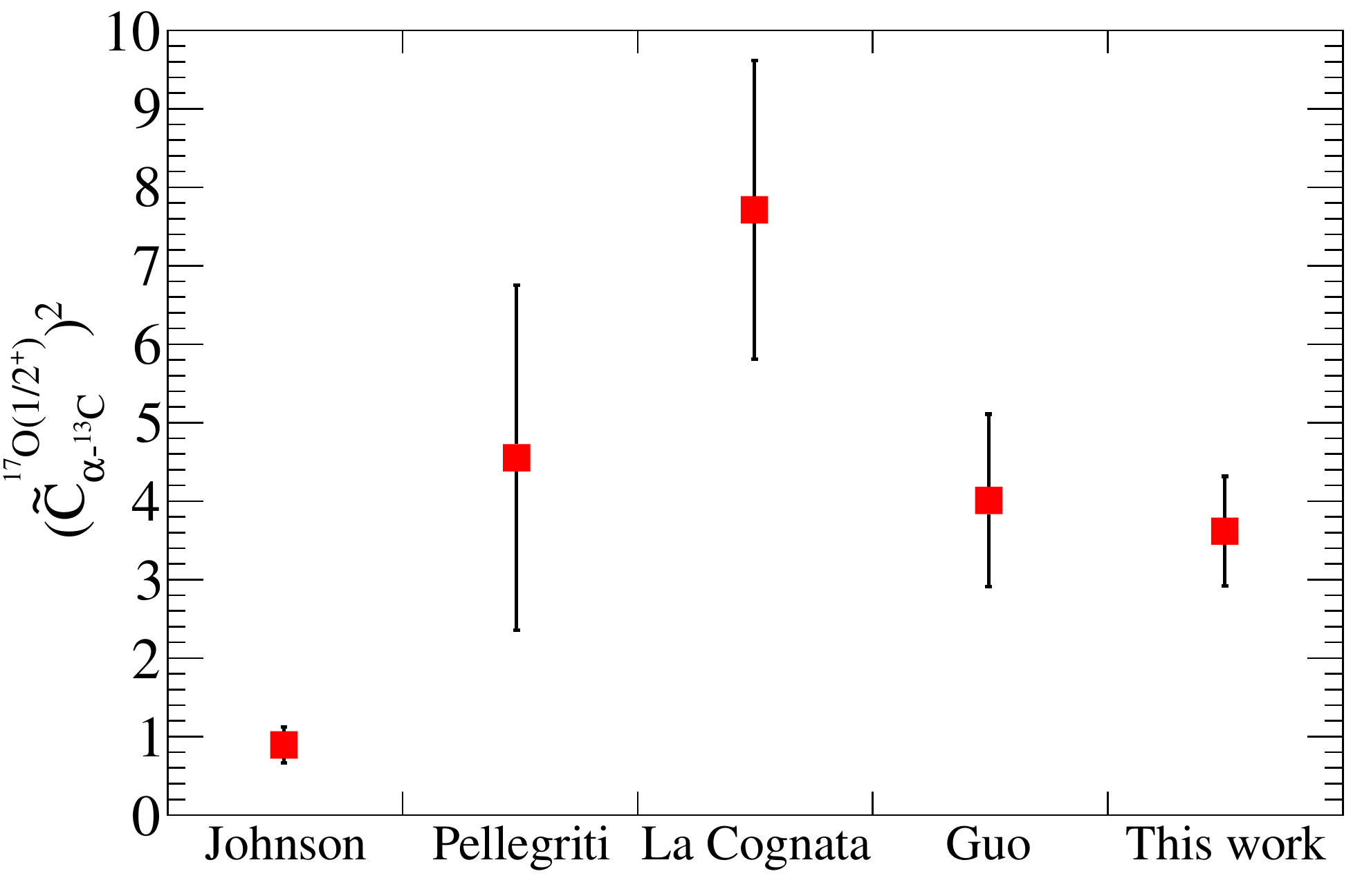}\end{center}
\caption{\label{fig:ANC_results} Squared Coulomb modified ANC value obtained in this work, for
the 1/2$^+$ at excitation energy of 6.356 MeV in $^{17}$O, compared to previous results from 
\citet{John06}, \citet{Pell08}, \citet{LaCo13} and \citet{Guo12}.}
\end{figure}

The squared Coulomb modified ANC that we determined in this work for the 1/2$^+$ state at 6.356 MeV in $^{17}$O is
$(\tilde C^{^{17}\text{O}(1/2+)}_{\alpha-^{13}\text{C}})^2=3.6\pm0.7\hspace{0.2cm}\text{fm}^{-1}$.
The ANCs reported in 
\cite{John06,Pell08,LaCo13,Guo12} and the value obtained in this work 
are shown in Figure \ref{fig:ANC_results}. In Table \ref{tab:ANC_summary} a summary 
of the results for the squared Coulomb modified ANC values and spectroscopic factors S$_\alpha$ are shown.
The ANC value of this work is within the error bars of the value found in \cite{Pell08,Guo12}.
However, it is smaller than the value found using the Trojan Horse method in Ref. \cite{LaCo13}. 
Compared to the ANC value obtained in the previous sub-Coulomb measurement in Ref. \cite{John06} the ANC obtained in
this work is about 4 times larger. The reason for this discrepancy is explained below.

The cross section measured at the effective energy of 7.72 MeV in this work is compared to the cross section from
\citet{John06} at 7.81 MeV in Fig. \ref{fig:ang_dist_17O}. The cross section measured in \cite{John06}
is about a factor of 3 smaller than the cross section measured in the present work. The main reason for this difference
is the target deterioration effect that was not taken into account in \citet{John06}. As mentioned above, 
the target thickening due to material buildup was observed in this experiment, causing the beam energy in the middle 
of the target to decrease. The effect was mitigated in this work by frequent target change and also by monitoring 
of the target condition using elastic scattering of $^6$Li beam. This was not done in the previous experiment \cite{John06}. 
Therefore, the beam energy in the middle of the target was decreasing during measurements that used the same target 
over extended period of time in Ref. \cite{John06}. As a result the measured cross section is significantly lower than it should have been. 
One of the characteristic features of sub-Coulomb transfer reactions (unlike the reactions performed at higher energies)
is strong dependence of the reaction cross section on the energy of the beam. The lower reaction cross section measured 
in \cite{John06} naturally lead to the smaller ANC.

\begin{table}
\centering
\caption{\label{tab:ANC_summary} Summary of the previous and current results for the
squared Coulomb modified ANC and spectroscopic factor $S_\alpha$ for the 1/2$^+$ sub-threshold resonance at
excitation energy of 6.356 MeV in $^{17}$O.}
\tabcolsep=0.45cm
\begin{tabular}{ccc}
\hline
\hline
$\left(\tilde C^{^{17}\text{O}(1/2+)}_{\alpha-^{13}\text{C}}\right)^2$  (fm$^{-1}$)& $S_\alpha$& Ref. \tabularnewline
\hline
	-	& 0.01	& \cite{Kubo03} \tabularnewline
0.89$\pm$0.23	&	-& \cite{John06}\tabularnewline
	-	& 0.36-0.40	& \cite{Keel03}	\tabularnewline
4.5$\pm$2.2	& 0.29$\pm$0.11	& \cite{Pell08}	\tabularnewline
7.7$\pm0.3_{\text{stat}}$$_{-1.5\hspace{0.05cm}\text{norm}}^{+1.6}$ & -	& \cite{LaCo13} \tabularnewline
4.0$\pm$1.1 & 0.37$\pm$0.12	& \cite{Guo12} \tabularnewline

3.6$\pm$0.7 & 	-   & This work	 \tabularnewline
 \hline
 \hline
\end{tabular} 
\end{table}

The importance of accurate knowledge of the $^{13}$C($\alpha,n)^{16}$O reaction rate has been emphasized recently in Refs. 
\cite{Guo12,LaCo13}.
This report provides a more precise and almost model independent value for the ANC of the 1/2$^+$ state at
6.356 MeV in $^{17}$O. The ANC for this state is the most important parameter that determines the $^{13}$C($\alpha,n)^{16}$O 
reaction rate at energies relevant for the s-process in AGB stars. This calls for a new complete R-matrix analysis of
all available experimental data that leads to $^{17}$O (of the type performed in Ref. \cite{Heil08}) to determine the new 
``recommended''  $^{13}$C($\alpha,n)^{16}$O reaction rate. We expect that the new, small uncertainty for the 1/2$^+$ at
6.356 MeV ANC value and also the new $^{13}$C+$\alpha$ elastic scattering data \cite{Mynb14} at low energies will provide 
reliable reaction rate constraints 
that will be adequate for modern models of AGB stars. While this result does not eliminate the need for direct
measurements of the $^{13}$C($\alpha,n)^{16}$O reaction cross section at as low an energy as possible, it reduces the nuclear
physics uncertainties of the s-process dramatically. It is important to point out that this result also resolves the
ambiguities arising from previous experimental results.

\section{Summary} 
In summary, we used the direct $\alpha$-transfer reaction $^6$Li$(^{13}$C$,d)^{17}$O at a sub-Coulomb energy to 
extract the $\alpha$ ANC for the 1/2$^+$ state at 6.356 MeV in $^{17}$O. This parameter is the major source of
uncertainty for the astrophysically important $^{13}$C$(\alpha,n)^{16}$O reaction rate at temperatures relevant for 
the s-process in AGB stars ($<$100 MK). The Coulomb modified squared ANC was determined to be 
$(\tilde C^{^{17}\text{O}(1/2+)}_{\alpha-^{13}\text{C}})^2=3.6\pm0.7\hspace{0.2cm}\text{fm}^{-1}$.
This is the most precise value to date but in good agreement with the results of Refs. \cite{Pell08,Guo12}. The main value of 
this work is that the discrepancy between the present results obtained by $\alpha$-transfer reactions at higher energy and the
sub-Coulomb energies is now removed. Both give similar values but the advantage of sub-Coulomb transfer is
that this technique is much less model dependent. The discrepancy (although a much smaller one than before)
still remains between the THM measurements and the sub-Coulomb ANC results. It is important to investigate 
the source of this discrepancy further in order to increase the reliability of both indirect methods, that
promise to be important tools for nuclear astrophysics. The more accurate ANC for the 1/2$^+$ at 6.356 
MeV state in $^{17}$O and the new low energy $\alpha$+$^{13}$C elastic scattering data \cite{Mynb14} can now be used 
to impose tighter constraints than before on the $^{13}$C$(\alpha,n)^{16}$O reaction rate.

\begin{acknowledgments}
The authors acknowledge the financial support provided by the National Science Foundation (USA)
under grant No. PHY-456463. G.V.R and E.K. acknowledge that this material is based upon their work supported
by the U.S. Department of Energy, Office of Science, Office of Nuclear Science, under Award 
No. DE-FG02-93ER40773. G.V.R also acknowledges the financial support of the Welch Foundation (USA) (Grant No.: A-1853).
\end{acknowledgments}

\bibliography{myrefs}

\end{document}